\documentclass[aps,prd,showpacs,reprint,nofootinbib,showkeys]{revtex4-1}

\usepackage{amsmath}
\usepackage{amsfonts}
\usepackage{amssymb}
\usepackage{graphicx}
\usepackage{hyperref}
\usepackage{braket}
\usepackage{enumitem}
\usepackage{bbm}
\usepackage{units}

\DeclareMathOperator{\arcosh}{arcosh}

\def\sl3c{\text{SL}(3,\mathbb{C})}
\def\su3{\text{SU(3)}}
\newcommand{\e}[1]{\times 10^{#1}}
\newcommand{\eitheta}{\braket{e^{i\theta}}}

\graphicspath{{plots/}}

\begin{document}  % for PhysRev

\title{Anatomy of a strong residual sign problem on the thimbles}

\author{Jacques Bloch}
\email{jacques.bloch@ur.de}
\affiliation{Institute for Theoretical Physics, University of Regensburg, 93040
  Regensburg, Germany}

\date{August 2, 2018}
%\date{\today}

\begin{abstract}
Using a simple Gaussian-like Ansatz for the phase distribution of a theory with a complex action, we show how the thimble integration for the average phase factor can be plagued by a strong residual sign problem when the phase of the complex integration measure conspires with the constant phase of the integrand along the thimble. This strong sign problem prohibits the accurate computation of the average phase factor when it becomes exponentially small, and causes a strong sensitivity to the parameters describing the phase distribution.

\hfill\textit{To the memory of Mike Pennington}
\end{abstract}

%\keywords{Sign Problem, Lattice QCD, Quark Chemical Potential}

\maketitle

\section{Introduction}

In lattice simulations of Quantum Chromodynamics (QCD) at nonzero chemical potential the action in the partition function is generically complex such that standard importance sampling Monte Carlo methods are unusable. One way to circumvent this problem is to apply the well known density of states method to this particular setting by splitting the complex action in its real and imaginary parts, and considering the density of the phase of the complex weights in the partition function generated by their magnitudes \cite{Gocksch:1988iz}. 

To apply the density of states method, the phase density $p(\theta)$ is measured explicitly in the phase quenched ensemble, and then integrated over to compute the average phase factor 
\begin{align}
\eitheta = \int d\theta \, p(\theta) \, e^{i\theta} .
\label{exp}
\end{align}
This can then be used to access thermodynamical observables \cite{Garron:2017fta} as the full and phase quenched partition function are related by $Z_\text{full}=\eitheta Z_\text{pq}$.
It is well known that for growing chemical potential, $\eitheta$ is exponentially small in the volume of the simulated system, and its computation is plagued by a strong sign problem \cite{deForcrand:2010ys}.
If $\eitheta$ is computed from the oscillatory integral \eqref{exp}, its accurate determination requires a very precise knowledge of the phase distribution $p(\theta)$, which might be obtained with the LLR method \cite{Langfeld:2014nta}. To further improve the accuracy and stability of the integration, a judicious fit to the measured phase distribution is performed before integrating the phase factor \cite{Garron:2017fta}.

Herein we present a simple example illustrating that the average phase factor obtained from such a fit is neither necessarily stable under slight variations of the fit parameters, nor can it be computed accurately when the sign problem becomes too strong. To compute this oscillatory integral we opted to use the thimble integration, which is often believed to reduce the sign problem and make it manageable. One of our motivations was to investigate the mechanism by which the thimble integration yields exponentially small values for $\eitheta$.

When integrating along a thimble, the magnitude of the complex integrand falls off in a Gaussian like manner at either side of the saddle point. Nevertheless, there is a potential sign problem due to the \textit{residual phase} along the thimble, which is caused by the phase of the complex measure along the integration path and the constant phase of the integrand on the thimble. Even though it is often claimed in the literature that the sign problem caused by this residual phase is most probably negligible \cite{Cristoforetti:2012su}, this is not true for the simple, physically motivated example discussed in this paper, as the phase of the complex measure can conspire with the constant nonzero phase of the integrand along the thimble to cause a sign problem that can even be maximally strong for physically relevant parameter values.

Although we applied the thimble integration to the one-dimensional oscillatory integral \eqref{exp} occurring in the density of states method,  the knowledge about the residual sign problem could have a wider scope as the Lefschetz thimbles, which are paths of steepest descent, are also intensively being explored to resolve the sign problem in the high dimensional integration occurring in the partition function of lattice QCD at nonzero chemical potential \cite{Cristoforetti:2012su,Cristoforetti:2013wha,Mukherjee:2013aga,DiRenzo:2017igr}. 

In section \ref{sec:phasedistrib} we will describe the choice of the phase distribution $p(\theta)$ in \eqref{exp}. In section \ref{sec:thimbles} we will introduce the basic elements needed to perform the thimble integration, and in section \ref{sec:results} we will give the results and discuss the residual sign problem. Finally, we close with some conclusions in \ref{sec:conclusions}.

\section{Phase distribution}
\label{sec:phasedistrib}

In lattice simulation of QCD at nonzero chemical potential the weights in the partition function are generically complex due to the fermion determinant. 
The distribution $p(\theta)$ represents the probability density of the phase of these complex weights in the phase quenched ensemble \cite{Gocksch:1988iz}, which is generated with the magnitude of the complex weights.

As $\theta$ is a complex phase it would be natural to assume $\theta \in (-\pi,+\pi]$. When the sign problem is large, the distribution nears a uniform distribution over this interval, and the exponentially small value for $\eitheta$  arises from tiny corrections to this uniform distribution, which cannot be determined accurately enough to allow for a useful determination of $\braket{e^{i\theta}}$. 

To improve upon this, it was suggested to consider an \textit{extensive phase}, which is defined such that $\theta$ is no longer bounded  and branch cut discontinuities are avoided \cite{Ejiri:2007ga,Nakagawa:2011eu}. 
Claim is that as the physical volume of the system gets larger, the distribution of the extended phase converges to a Gaussian distribution such that a pure Gaussian Ansatz would suffice to compute $\braket{e^{i\theta}}$
and perform phenomenology at nonzero density \cite{Ejiri:2007ga,Ejiri:2007tk,Ejiri:2008xt,Ejiri:2010vm,Nakagawa:2011eu,Ejiri:2012wp}. 
The argumentation also involves the cumulant expansion for $\braket{e^{i\theta}}$, which, if it converges, is always real and positive, and whose leading term corresponds to the Gaussian result. 

The Gaussian Ansatz was however questioned by the observation that higher order corrections in the cumulant expansion, which involve delicate volume cancellations, could invalidate the Gaussian value of $\braket{e^{i\theta}}$ \cite{Greensite:2013vza,Greensite:2013gya,Greensite:2013ska}. The simple example presented below is very much supporting the latter argument, as we found that the cumulant expansion converges extremely slowly for the simple Gaussian-like distribution \eqref{fit} when the sign problem becomes strong, and that higher order terms can indeed make $\eitheta$ orders of magnitudes smaller than its naive Gaussian value. The detailed discussion of the convergence of the cumulant expansion will be discussed elsewhere \cite{Bloch:2018}, as we will presently focus on the analysis of the thimble integration.

The distribution $p(\theta)$ of the extended phase is generically described by an exponential of an even polynomial in $\theta$, as is discussed in the context of Monte Carlo simulations of QCD at finite density \cite{Greensite:2013gya,Greensite:2013ska,Garron:2017fta}. 
Herein we will consider the simplest extension of the Gaussian distribution within that framework, namely an exponential of a quartic polynomial,
\begin{align}
p(\theta) = N \exp\left[-\frac{\theta^2}{2\sigma^2}\left(1+a\frac{\theta^2}{\sigma^2}\right)\right]  
\label{fit}
\end{align}
with normalization factor $N = 2\sqrt{a}/(\sigma e^{\kappa}K_{1/4}(\kappa))$, where $\kappa= 1/(16a)$, $K_{1/4}$ is a modified Bessel function of fractional order, and $a\geq0$. 

This particular functional form for $p(\theta)$ is also suggested by detailed studies of the phase distribution in Monte Carlo simulations of a random matrix theory (RMT) that models some important properties of QCD at nonzero chemical potential \cite{Osborn:2004rf,Stephanov:1996ki} and which will be reported elsewhere \cite{Bloch:2018}. 
The quartic term is dictated by the tails of the measured phase distributions, which are slightly narrower than those of a normal distribution. The parameters $\sigma$ and $a$ can unambiguously be extracted from the second and fourth moments of the phase distribution measured in the Monte Carlo simulations. 

Although the analysis performed in this paper is valid for any value of $\sigma$ in \eqref{fit}, the results will all be given for $\sigma=4.2$. We will investigate the behavior of $\eitheta$ for small values of $a$, where the distribution is very close to normal.
This choice of parameter values stems from the RMT simulations for matrix sizes where the sign problem is strong \cite{Bloch:2018}. 

\begin{figure}
\centering
\includegraphics{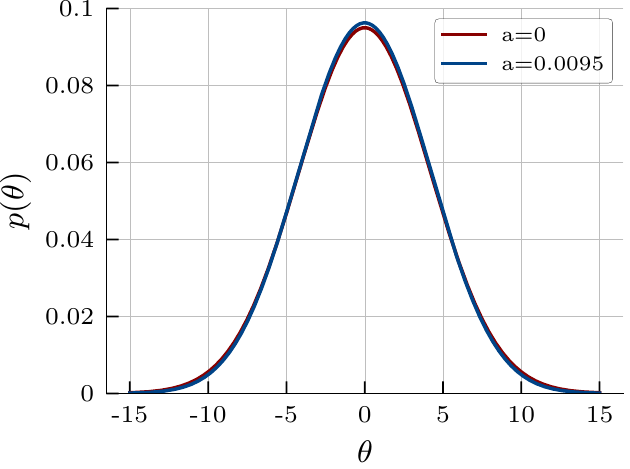}
\caption{Distribution $p(\theta)$ of \eqref{fit} for $\sigma=4.2$ and $a=0.0095$ (blue) compared to the Gaussian distribution ($a=0$) (red). 
}
\label{fig:distrib}
\end{figure}

\begin{figure}
\centering
\includegraphics{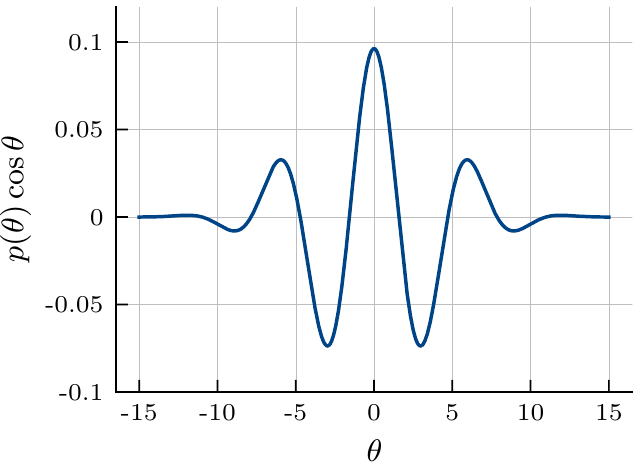}
\caption{Real part $p(\theta)\cos\theta$ of the oscillating integrand in \eqref{exp} with phase distribution \eqref{fit} for $\sigma=4.2$ and $a=0.0095$.}
\label{fig:integrand}
\end{figure}

The phase distribution \eqref{fit} is illustrated in Fig.~\ref{fig:distrib} for $\sigma=4.2$ and $a=0.0095$. For such small $a$ the distribution is almost undistinguishable from a Gaussian, as can be seen in the figure. 
In Fig.~\ref{fig:integrand} we show the real part of the oscillating integrand in \eqref{exp} obtained for this distribution.

\begin{figure}
\centering
\includegraphics{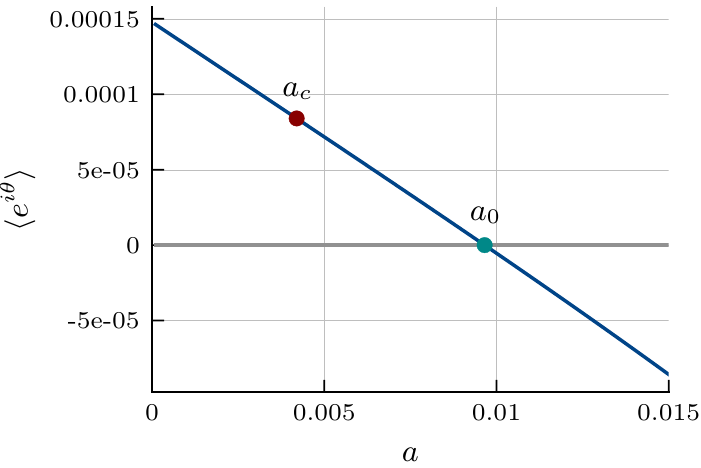}
\caption{Integral $\Braket{e^{i\theta}}$ as a function of $a$ with phase distribution \eqref{fit} for $\sigma=4.2$. The zero crossing is denoted by $a_0$. The point $a_c$ indicates the transition from the single to double-thimble region, where the residual sign problem sets in. }
\label{fig:avgcos}
\end{figure}

Although the main focus of this paper will be the computation of \eqref{exp} using Lefschetz thimbles, in particular in the region where the sign problem is strong, we can also compute this one-dimensional integral using standard numerical quadrature routines, as long as the sign problem remains amenable to such methods. The results for $\Braket{e^{i\theta}}$ as a function of $a$ are presented in Fig.~\ref{fig:avgcos} for $\sigma=4.2$. When $a=0$ we recover the Gaussian result, where the average phase factor can be computed analytically,
\begin{align}
\Braket{e^{i\theta}}_\text{Gauss} = e^{-\sigma^2/2} .
\label{expGauss}
\end{align}
As $a$ increases, the average phase factor decreases and has a zero crossing at $a_0 \approx0.00965632$. The region $a\in[0,a_0)$ is especially important in the context of physical simulations with a complex action as $\braket{e^{i\theta}}$ is known to be positive but exponentially small in the volume. In the remainder of this paper we will investigate how such small values can arise in the thimble framework.

\section{Thimble analysis}
\label{sec:thimbles}

In the thimble formulation \cite{Witten:2010cx}, the original integral \eqref{exp} is rewritten as
\begin{align}
\braket{e^{i\theta}} = \int d\theta \, p(\theta) \, e^{i\theta} 
= \sum_{{\cal T}\in\Omega} I_{\cal T} ,
\label{thimbles}
\end{align}
where $\Omega$ is the set of relevant thimbles in the complex plane that contribute to the integral. Thimbles are trajectories of constant phase going through saddle points of the integrand. The integral on a thimble ${\cal T}$ is 
\begin{align}
I_{\cal T} &= \int_{\cal T} dz \, f(z) = \int_{\cal T} ds \, f(z(s)) \, \frac{dz}{ds} ,
\end{align}
where we changed notation from $\theta$ to $z$, to emphasize that the variable has been complexified, and parametrized the thimble by its arc length $s$.
The complex measure along the thimble can be rewritten as $dz = ds \, e^{i\eta(s)}$, such that
\begin{align}
I_{\cal T} &=  \int_{\cal T} ds \, f(z(s)) \, e^{i\eta(s)} .
\end{align}
The phase factor $e^{i\eta(s)}$ is the Jacobian of the arc length parametrization of the thimble, where $\eta(s)$ is the angle of the tangential to the thimble.
 
As thimbles are trajectories of constant phase $\phi$, we can write $f(z(s)) = r(s) \, e^{i\phi}$ with $r(s)=|f(z(s))|$ such that
\begin{align}
I_{\cal T} &=  \int_{\cal T} ds  \, r(s) e^{i\left(\phi+\eta(s)\right)} .
\label{IT}
\end{align}
In this form the equation will be most useful to investigate the strong residual sign problem.

Let us first briefly consider a pure Gaussian distribution in \eqref{exp}. It is well known how the thimble construction trivially solves the sign problem in this case. The single saddle point, given by the zero of the derivative of the action, is located at $z_0=i\sigma^2$ and the thimble is parallel with the real axis, with constant phase $\phi=0$. The integrand, which was strongly oscillating on the real axis, is now replaced by a Gaussian integrand on the thimble. 
The integral value becomes exponentially small, while avoiding a sign problem, because the function value in the saddle point becomes exponentially small when the integration contour is pushed up in the complex plain. 

A salient feature of the Gaussian distribution is that the average phase factor \eqref{expGauss} is always positive and not very sensitive to small changes in the width of the distribution. We will see that this is no longer true when generalizing the phase distribution to \eqref{fit}, as a very different thimble mechanism is at work close to the zero crossing $a_0$.

%Before investigating the thimble structure for the phase distribution \eqref{fit} in detail, we can already remark that a very different thimble mechanism must be at work close to the zero crossing $a_0$ (see Fig.~\ref{fig:avgcos}), as infinitesimally small values for $\eitheta$ cannot merely be obtained by considering the function value in the saddle point.
%, which will only slightly change for small changes in $a$. 

To determine the saddle points and thimble trajectories we rewrite the integrand as $f(z)=e^{-S(z)}$ with complex action $S$. For the integrand in \eqref{exp} with phase distribution \eqref{fit} the complex action is
\begin{align}
S(z) = \frac{z^2}{2\sigma^2} + \frac{a z^4}{2\sigma^4} - iz - \log N
\label{action}
\end{align}
with saddle point equation
\begin{align}
\frac{\partial S}{\partial z} =  \frac{z}{\sigma^2} + \frac{2az^3}{\sigma^4} - i = 0 .
\end{align}
After rewriting the saddle point solutions as $z=it$, the equation becomes a cubic equation in $t$,
\begin{align}
t^3 + p\,t + q = 0 
\label{cubic}
\end{align}
with real coefficients
\begin{align}
p = -\frac{\sigma^2}{2a} ,\quad
q = \frac{\sigma^4}{2a}
.
\label{pq}
\end{align}
Depending on the value of the discriminant 
\begin{align}
\Delta=-4p^3-27q^2 ,
\end{align}
this equation has either three real solutions if $\Delta\geq0$ or one real and two complex conjugate solutions if $\Delta<0$. After substituting \eqref{pq} in $\Delta$, we find that the discriminant is zero when $a=a_c$ with critical value
\begin{align}
a_c = \frac{2}{27\sigma^2}.
\label{transition}
\end{align}
When $a\leq a_c$ the quartic term in \eqref{fit} is small and the distribution becomes more Gaussian-like. In this case, $\Delta \geq 0$ and \eqref{cubic} has three real roots \cite{Dickson1914}
\begin{align}
t_k = 2\sqrt{-\frac{p}{3}}\cos\left[\frac13\arccos\left(\frac32\frac{q}{p}\sqrt{-\frac{3}{p}}\right)-\frac{2\pi k}{3}\right] ,
\label{realsol}
\end{align}
with $k=0,1,2$. The three saddle points $z_k=it_k$ are located on the imaginary axis. 

Although the solutions \eqref{realsol} can be analytically continued to $a>a_c$, where $\Delta<0$, it is more revealing in this case to write the real solution $t_0$ as \cite{Holmes2002}
\begin{align}
t_0 = -2\sqrt{-\frac{p}{3}}\cosh\left(\frac13\arcosh\left(-\frac{3}{2}\frac{q}{p}\sqrt{-\frac{3}{p}}\right)\right)
\end{align}
and the two complex conjugate solutions $t_\pm$ as solutions of the remainder quadratic equation, 
 \begin{align}
t^2 + t_0 t + (p+t_0)^2 = 0 ,
\end{align}
obtained by dividing \eqref{cubic} algebraically by $t-t_0$. Its complex conjugate solutions are given by
\begin{align}
t_{\pm} = -\frac{t_0}{2} \pm \frac{i}{2} \sqrt{4p+3t_0^2} .
\label{tpm}
\end{align}
There are thus three saddle points: $z_0=it_0$ on the imaginary axis, and a complex pair $(z,-z^*)=(it_+,it_-)$, located symmetrically left and right of the imaginary axis. 

Note that the saddle points are explicit functions of the parameters $\sigma$ and $a$ of the phase distribution \eqref{fit}.

Once the saddle points are known, a further analysis is performed to determine which thimbles are relevant to the thimble integration.

For $a \leq a_c$ the thimble structure is quite similar to that of the Gaussian distribution and only one thimble, going through the saddle point $z_1=it_1$ of \eqref{realsol}, contributes to the integral. 
The constant phase $\phi$ along the thimble is zero, as can either be computed explicitly from the action in the imaginary saddle point, or can be deduced from the fact that the integral \eqref{exp} is known to be real.

On the other hand, for $a>a_c$ the thimble through the imaginary saddle point does not contribute to the integral, rather, the thimble integration is now given by the sum of the two thimbles ${\cal T}_-$ and ${\cal T}_+$ that are mirrored about the imaginary axis and go through the $(z,-z^*)$ pair of saddle points corresponding to \eqref{tpm}. Crucial is that $\phi$ is not constrained to zero on the mirrored thimbles, and its nonzero value will cause the strong residual sign problem on the thimbles, as will be discussed in more detail below.

\begingroup
\squeezetable
\begin{table}[b]
\centering
\caption{Summary of results for the thimble analysis of the integral \eqref{exp} with distribution \eqref{fit} for $\sigma=4.2$ and varying $a$, showing the relevant saddle points, constant phase and average phase factor. 
}
\label{tab:results}
\begin{tabular}{|c|c|c|c|c|c|}
\hline
& $a$ & $z_0$ & $\phi$ & $\braket{e^{i\theta}}$\\
\hline
Gauss & $0$ & $i\,17.64$ & 0 & $1.477\e{-4}$ \\
\hline
single- & 0.001 & $i\,18.3393$ & 0 & $1.326\times 10^{-4}$   \\
\hfill thimble & 0.004 & $i\,23.6046$ & 0 & $8.698\times 10^{-5}$ \\
\hline
double- & 0.00425 & $\pm 1.66704 + i\, 26.319$ & 0.0066 & $8.318\times 10^{-5}$ \\
\hfill thimble & 0.009 & $\pm 9.85516 +i\, 18.9484$ & 2.098 & $1.022\e{-5}$ \\
& 0.0093 & $\pm 9.93506 +i\, 18.6824$ & 2.190 & $5.556\e{-6}$ \\
& 0.0095 & $\pm 9.98268 +i\, 18.5119$ & 2.249 & $2.439\e{-6}$ \\
& 0.00965 & $\pm 10.0157 +i\, 18.3875$ & 2.292 & $9.869\e{-8}$ \\
& 0.009656 & $\pm 10.0170 +i\, 18.3826$ & 2.293 & $5.022\e{-9}$ \\
& 0.0096563 & $\pm 10.0171 +i\, 18.3824$ & 2.293 & $3.383\e{-10}$ \\
& 0.00965632 & $\pm 10.0171 +i\, 18.3824$ & 2.293 & $2.610\e{-11}$ \\
\hline
\end{tabular}
\end{table}
\endgroup

\section{Results}
\label{sec:results}

The aim is to understand how exponentially small integral values arise in the thimble framework, and to investigate the impact of small variations in $a$ on $\braket{e^{i\theta}}$. Some thimble properties that will be discussed throughout this section are summarized in Table~\ref{tab:results}. All results in this section were obtained for $\sigma=4.2$.

We first look at the constant phase along the relevant thimbles. For this, we substitute the value of the saddle point on the relevant thimble in the action \eqref{action} and set $\phi=-S_I$, with $iS_I$ the imaginary part of the action.

\begin{figure}[t]
\centering
\includegraphics{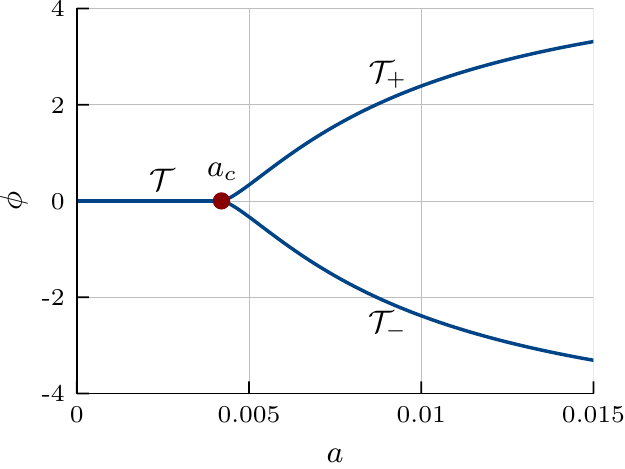}
\caption{Constant phase $\phi$ on the thimbles as a function of $a$ in \eqref{fit} for $\sigma=4.2$. At $a=a_c$ the system changes from single to double-thimble mode and $\phi$ becomes nonzero.}
\label{fig:constphase}
\end{figure}

For $a\leq a_c$ the saddle point is purely imaginary and the constant phase is zero. 
For $a>a_c$ the action for the $(z,-z^*)$ pair of saddle points satisfies $S(-z^*)=[S(z)]^*$, such that the constant phase $\phi$ has opposite values on both thimbles. The constant phase $\phi$ is shown as a function of $a$ in Fig.~\ref{fig:constphase}. The phase is zero for $a\leq a_c$ and becomes nonzero when $a>a_c$, increasing steadily as a function of $a$ in the region of interest. 

\begin{figure*}
\centering
\includegraphics{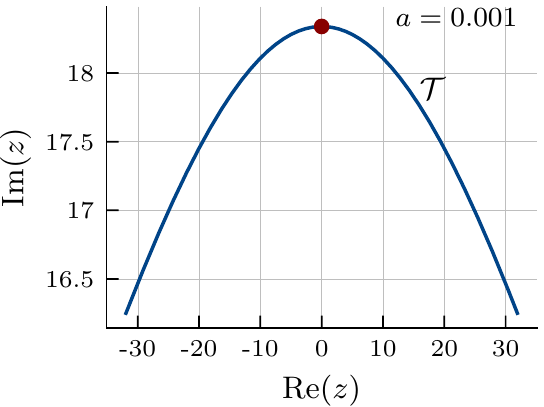}
\includegraphics{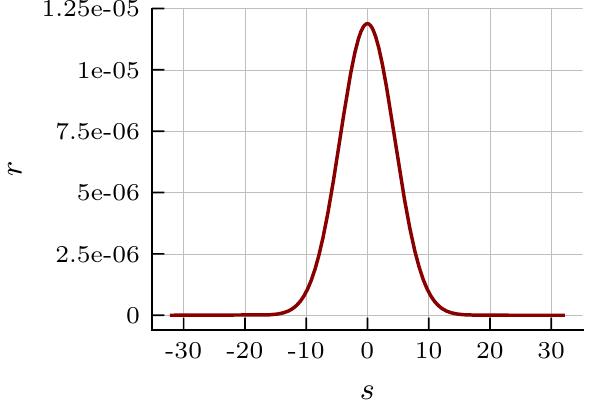}
\includegraphics{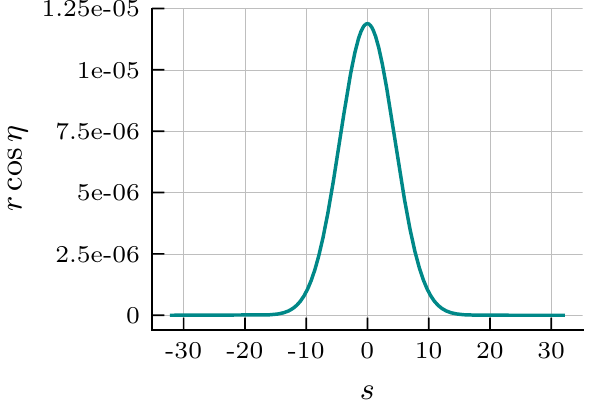}
\\
\includegraphics{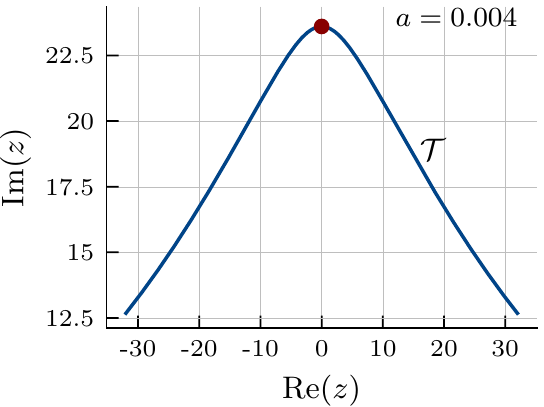}
\includegraphics{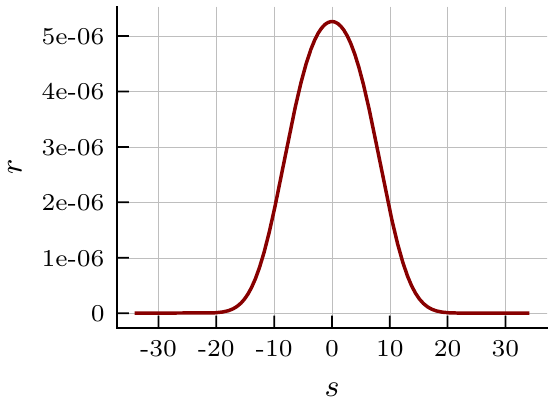}
\includegraphics{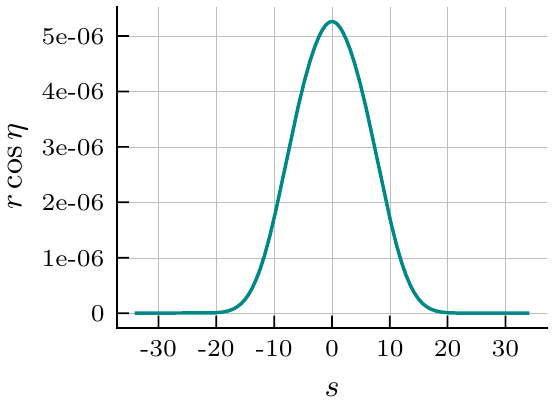}
\caption{Thimble analysis for $\sigma=4.2$ and $a=0.001$ (top) and $0.004$ (bottom). As $a<a_c$, the integral is given by a single-thimble. The left pane shows the thimble trajectory in the complex plane, with the saddle point represented by a red bullet. The middle pane shows the magnitude $r(s)$ and the right pane $r(s)\cos\eta(s)$ versus the arc length $s$ on the thimble (we explicitly set $\phi=0$). The saddle point is located at $s=0$.  }
\label{fig:thimbles1}
\end{figure*}

\begin{figure*}
\centering
\includegraphics{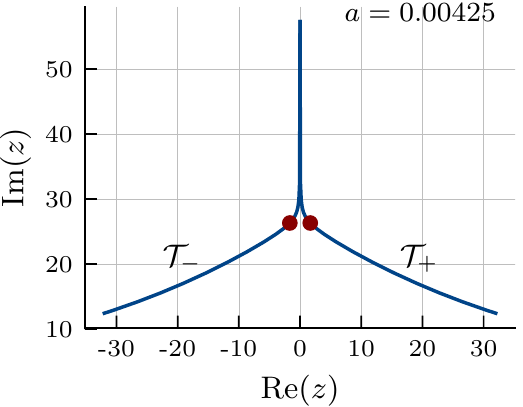}
\includegraphics{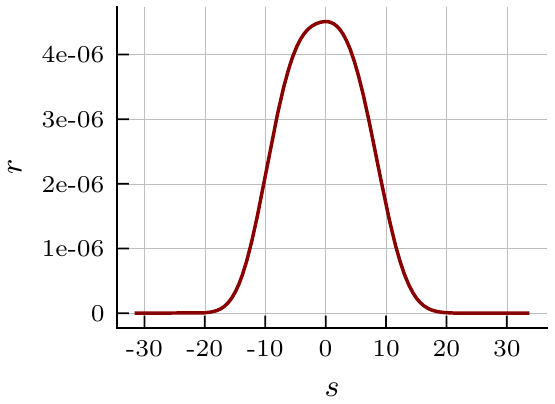}
\includegraphics{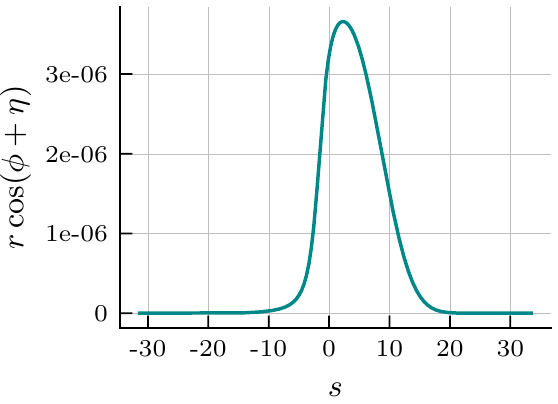}
\caption{Thimbles analysis for $\sigma=4.2$ and $a=0.00425$.
As $a>a_c$ the integral is given by two mirrored thimbles. Their trajectories in the complex plane are shown the left pane, with the saddle points represented by red bullets. The middle pane shows the magnitude $r(s)$ and the right pane $r(s)\cos(\phi+\eta(s))$ versus the arc length $s$ on the thimble ${\cal T}_+$. The saddle point is located at $s=0$.}
\label{fig:thimbles2}
\end{figure*}

\begin{figure*}
\centering
\includegraphics{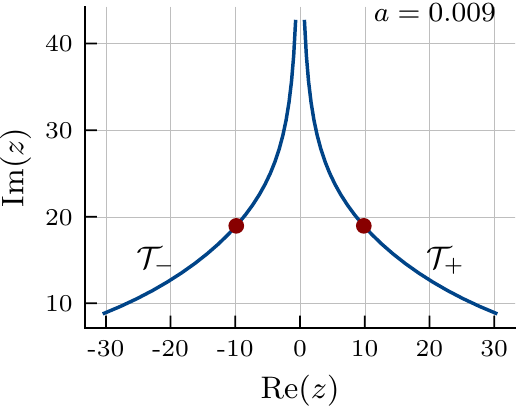}
\includegraphics{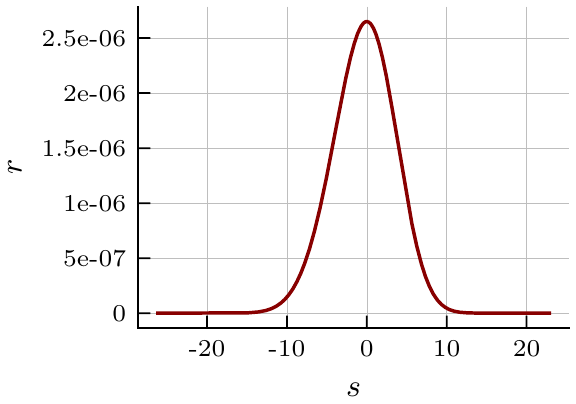}
\includegraphics{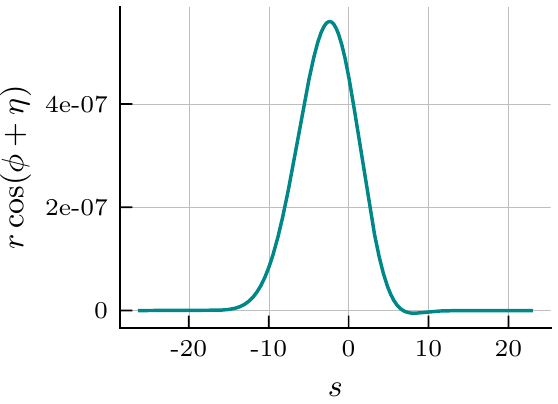}
\\
\includegraphics{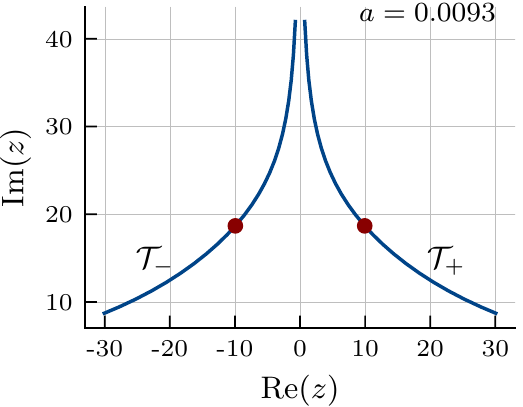}
\includegraphics{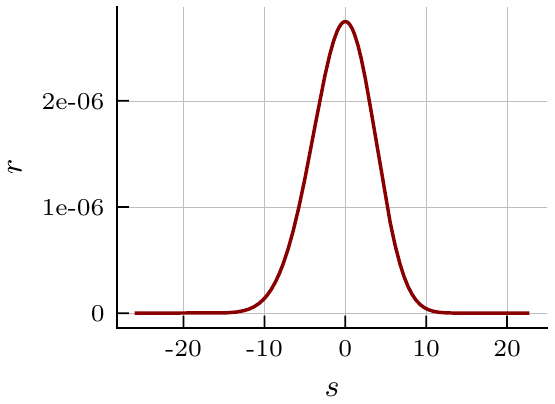}
\includegraphics{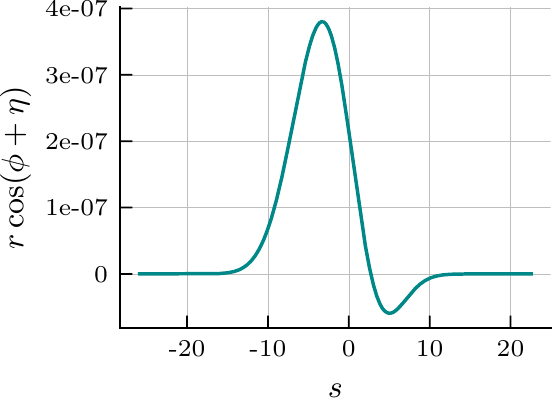}
\\
\includegraphics{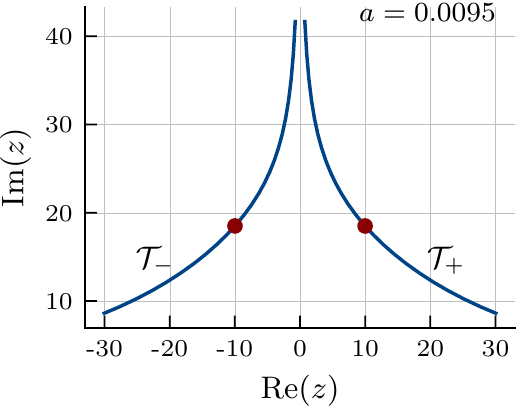}
\includegraphics{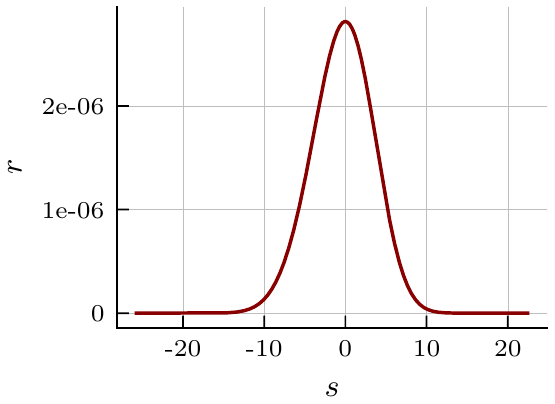}
\includegraphics{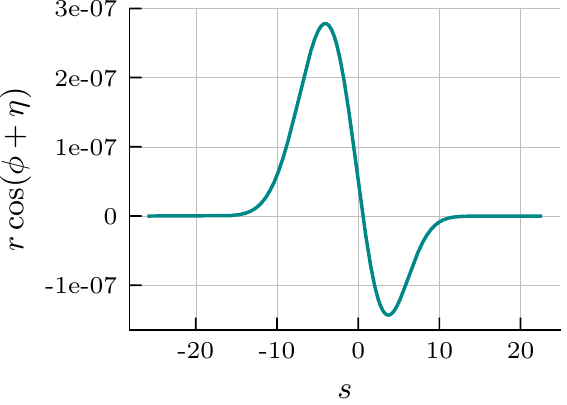}
\\
\includegraphics{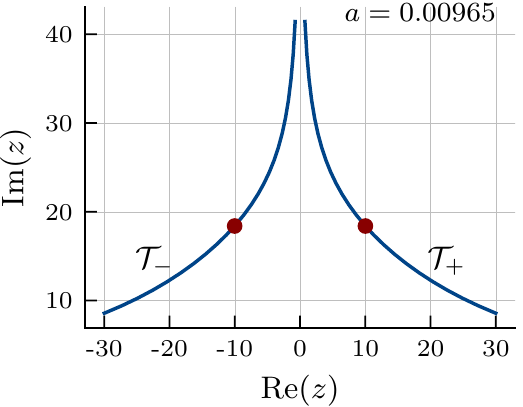}
\includegraphics{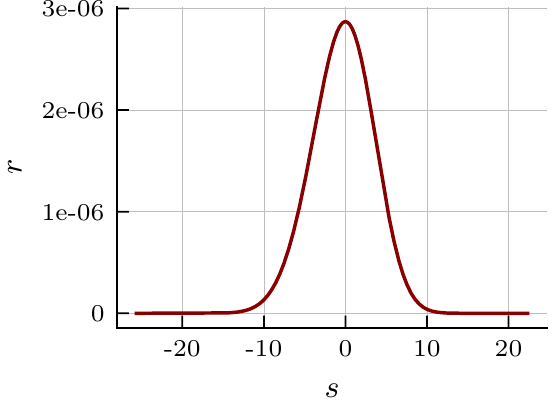}
\includegraphics{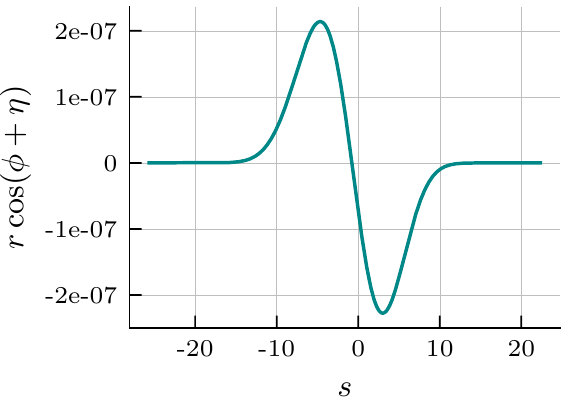}
\caption{Thimbles analysis for $\sigma=4.2$ and $a=0.009$, $0.0093$, $0.0095$, and $0.00965$ (from top to bottom).
As $a>a_c$ the integral is given by two mirrored thimbles.  The left pane shows both thimble trajectories in the complex plane, with the saddle points represented by the red bullets. The middle pane shows the magnitude $r(s)$ and the right pane $r(s)\cos(\phi+\eta(s))$ versus the arc length $s$ on the thimble ${\cal T}_+$. The saddle point is located at $s=0$.}
\label{fig:thimbles3}
\end{figure*}

When studying the thimble integration for various values of $a$ in Figs.~\ref{fig:thimbles1}-\ref{fig:thimbles3}, we will display three panes for each parameter value. From left to right:
\begin{enumerate}[label=(\Alph*)]
\item the path(s) of the relevant thimble(s) in the complex plane with their saddle point(s); 
\item the \textit{magnitude} $r(s)$ of $f$ as a function of the arc length parametrization $s$ along the thimble, see \eqref{IT}. As $\phi$ is constant along the thimble, the evolution of $r(s)$ tells us how fast the real and imaginary parts of $f$ fall off when moving away from the saddle point along the thimble. In practice, the thimbles were followed until the function dropped to $10^{-12}$ of its maximal value in the saddle point. 
\item the real part of $r(s) \, e^{i(\phi+\eta(s))}$, which is the actual function to be integrated over  after parametrizing the thimble by its arc length $s$, as in \eqref{IT}. This integrand includes the Jacobian $e^{i\eta(s)}$ of the parametrization, where $\eta(s)$ is the angle of the tangential to the thimble.
\end{enumerate}
In panes (B) and (C) the saddle point is chosen as origin of the arc length parametrization.
When two thimbles contribute to the integral, these two panes only show the functions along the ${\cal T}_+$ thimble, as those on ${\cal T}_-$ are related by complex conjugation.

We now analyze the thimble integration as function of the parameter $a$. For $\sigma=4.2$ the transition between the single-thimble and double-thimble regions happens at $a_c \approx 0.0042$, in accordance with \eqref{transition}. For $a=0$ the distribution \eqref{fit} is Gaussian. As $a$ increases, with $a\leq a_c$, the integral is still represented by a single relevant thimble with saddle point on the imaginary axis and constant phase $\phi=0$. 
This is illustrated in Fig.~\ref{fig:thimbles1}, where we show the thimble solutions for $a=0.001$ (top) and $a=0.004$ (bottom). The content of the three panes is as explained above in items (A)-(C). The thimble path in the complex plane is shown on the left in each row. In the middle we show the magnitude $r(s)$ of $f$ as a function of the arc length $s$, while on the right we show $r(s)\cos\eta(s)$, which is the real part of the product of $f(s)$ with the Jacobian $e^{i\eta(s)}$, as $\phi$ is zero below $a_c$. The latter is the actual integrand that yields $\braket{e^{i\theta}}$ after integration over $s$, see \eqref{IT}.
When performing the thimble integration, there is a small variation of the phase of the integrand along the thimble due to the Jacobian $e^{i\eta(s)}$. This fluctuation is however insignificant, as $\cos\eta(s)$ remains very close to one. This is confirmed by the observation that panes (B) and (C) of Fig.~\ref{fig:thimbles1} are almost indistinguishable.

\begin{figure*}
\centering
\includegraphics{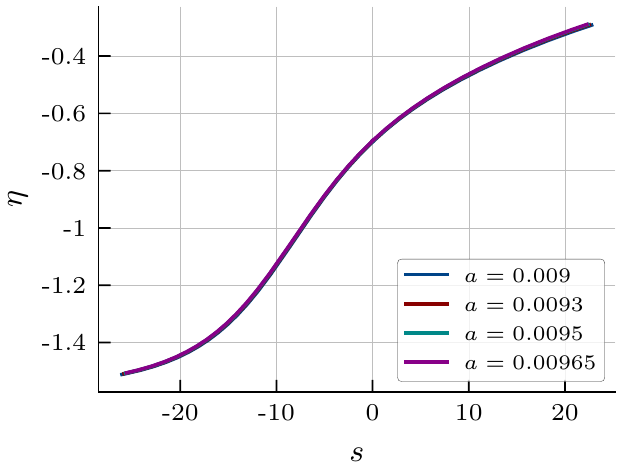}
\hspace{10mm}
\includegraphics{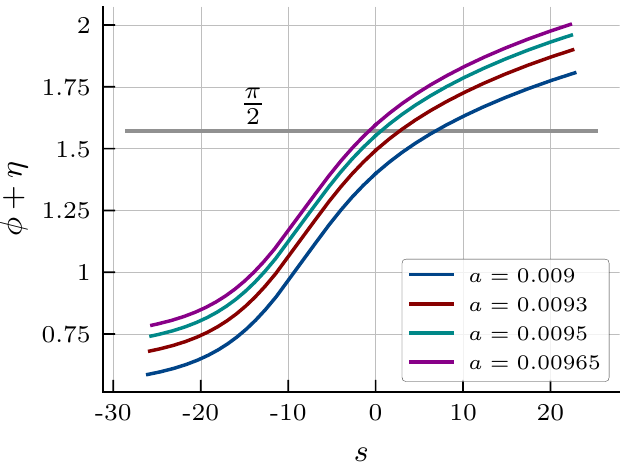}
\caption{Left: Phase $\eta(s)$ of the Jacobian as a function of the arc length $s$ along the thimble ${\cal T}_+$ for $\sigma=4.2$ and values of $a\in[0.009,a_0]$. The phase remains practically unaltered under such small variations of $a$, as the curves fall on top of one another. Right: Total phase $\phi+\eta(s)$ of the integrand as a function $s$ for the same parameter values. The variation of the constant phase $\phi$ with $a$ (see Fig.~\ref{fig:constphase}) is responsible for the separation of the total phases when $a$ is varied.}
\label{fig:eta}
\end{figure*}

When $a$ nears $a_c$ from below, two purely imaginary saddle points move toward each other along the imaginary axis, then merge when $a=a_c$, and move apart again as $a>a_c$, leaving the imaginary axis as a $(z,-z^*)$ pair of saddle points. The integral is then represented by the pair of thimbles ${\cal T}_-$ and ${\cal T}_+$, which are mirrored about the imaginary axis, as is illustrated in Fig.~\ref{fig:thimbles2}A for $a=0.00425$, which is just above $a_c$. From the action \eqref{action}, it is easy to verify that the contributions of the two thimbles to the integral are complex conjugate. 
Figure~\ref{fig:thimbles2} also illustrates how the $a\leq a_c$ region smoothly connects to the $a>a_c$ region: as the pair of saddle points emerges, the single-thimble splits into two mirrored thimbles. Let us focus on ${\cal T}_+$: the thimble is almost vertical for $s<0$, such that $\eta(s<0)\approx-\pi/2$.
As the constant phase $\phi\approx 0$ (see Table \ref{tab:results}), the real part of the phase factor $\cos(\phi+\eta(s)) \approx 0$ for negative $s$. 
This can be verified by comparing panels (B) and (C) in Fig.~\ref{fig:thimbles2}, showing the integrand on the thimble ${\cal T}_+$. Although the magnitude $r(s)$ in pane (B) is fairly symmetric about $s=0$, it clearly becomes asymmetric in pane (C) after multiplication with the phase factor $\cos(\phi+\eta(s))$ to compute the real part of the integrand, which almost vanishes for negative $s$. 
If we compare the thimble integration below $a_c$ and just above it, we see that the Gaussian-like curve on ${\cal T}$ in Fig.~\ref{fig:thimbles1}C turns into a sum of two half-Gaussians, one on ${\cal T}_+$, shown in Fig.~\ref{fig:thimbles2}C, and an identical one on ${\cal T}_-$.

We now further increase $a$, and zoom in on the region $a\in[0.009,a_0]$, with $a_0\approx0.00965632$. In Fig.~\ref{fig:thimbles3} we show the thimble analyses for the four parameter values $a=0.009$, $0.0093$, $0.0095$ and $0.00965$ (from top to bottom). The left plots show that the positions of the saddle points and the thimble paths do not change substantially under such small variations. This is confirmed by the locations of the saddle points given in Table~\ref{tab:results} and the plots of $\eta(s)$, which fall on top of each other in the left pane of Fig.~\ref{fig:eta}. As expected, the magnitude $r(s)$ in Fig.~\ref{fig:thimbles3}B is Gaussian-like and  scarcely changes under small variations in $a$.

When integrating along the thimble the magnitude $r(s)$ gets modulated with $\cos(\phi+\eta(s))$, and  
two observations should be made. Firstly, even though $\eta(s)$ is insensitive to small changes of $a$, it does evolve significantly as a function of $s$ along the thimble, as is shown in the left pane of Fig.~\ref{fig:eta}. Secondly, when increasing $a$ from $a_c$ to $a_0$, the constant phase $\phi$ on ${\cal T}_+$ increases from $\phi=0$ to $\phi\approx 2.3$.
More specifically, for $a$ varying from top to bottom in Fig.~\ref{fig:thimbles3}, the constant phase $\phi$ varies from 2.1 to 2.3, as can be read off from Table~\ref{tab:results} and Fig.~\ref{fig:constphase}.
These two facts are gathered in the right pane of Fig.~\ref{fig:eta} where we show the phase $\phi+\eta(s)$ for the values of $a$ investigated in Fig.~\ref{fig:thimbles3}. We observe that the intersection with $\pi/2$, where the cosine vanishes, shifts to values of $s$ well inside the relevant region of the thimble integration, and hence the real part of the integrand in \eqref{IT} oscillates, which generates a residual sign problem. 
This can clearly be observed in Fig.~\ref{fig:thimbles3}C, as the positive and negative contributions cancel more and more as the parameter $a$ is varied slightly from top to bottom in the figure and gets closer to $a_0$. This is also confirmed by the value of $\braket{e^{i\theta}}$ in Table~\ref{tab:results}, which decreases by seven orders of magnitude compared to the Gaussian result for the values of $a$ considered in the table. 
In fact, the residual sign problem can become arbitrarily strong as $a$ approaches $a_0$: the positive and negative contributions to the thimble integration will cancel to higher and higher precision as $\braket{e^{i\theta}}$ becomes smaller and smaller, and eventually goes to zero. 

At first sight, this may look like a rather artificial problem, considering that it is merely caused by the zero crossing of $\braket{e^{i\theta}}$ in Fig.~\ref{fig:avgcos}, and knowing that the sign problem will go away when increasing $a$ further and the integral turns negative.
However, this particular thimble mechanism is physically relevant as we know that $\eitheta$ is positive for physical systems with a complex action, as it can be written as a ratio of partition functions. Moreover, the parameter values $\sigma$ and $a$ of the phase distribution \eqref{fit}, obtained from Monte Carlo simulations of RMT at nonzero chemical potential, turn out to be such that $\braket{e^{i\theta}}$ is many orders of magnitudes smaller than its Gaussian value.
In this case the thimble integration does not solve the sign problem as the exponentially small values of $\eitheta$ are only obtained at the cost of a strong residual sign problem in the thimble integration itself.

\section{Conclusions}
\label{sec:conclusions}

The average phase factor $\eitheta$ is central in the density of states method used in simulations of physical systems with a complex action. As its measurement is plagued by a sign problem, it is sometimes improved by integrating over a judicious fit to the measured phase distribution. The hope being that $\eitheta$ could then be computed accurately and with little sensitivity to the fit parameters.

Although, the Gaussian distribution is a simple candidate for a fit function that allows for an accurate computation of $\eitheta$ and  is stable under small variations of its fit parameter, this paper has shown that tiny deviations from the Gaussian distribution can  readily ruin these properties.

We extended the Gaussian distribution with a quartic term in the exponential, which was suggested by simulations of random matrix theory at nonzero chemical potential. Higher order corrections were not considered as their contributions in these simulations were consistent with zero. Moreover, the simple extension considered in this paper suffices to show the deficiencies of the fitting Ansatz method. 

We performed a detailed analysis of the thimble integration to compute $\eitheta$ for parameter values that are physically relevant, i.e., with a quartic term that only slightly perturbs the leading order Gaussian part. We showed that there are fundamentally two regions depending on the strength of the quartic term: one where the integral is represented by a single thimble for which the constant phase of the integrand is zero, and another one where a pair of mirrored thimbles make up the thimble integration. In the latter case, the constant phase is nonzero, and, together with the phase of the Jacobian of the thimble parametrization, a strong residual sign problem arises, allowing $\eitheta$ to become exponentially small as $a$ nears $a_0$.

In Monte Carlo simulations of physical systems with a complex action, $\eitheta$ must be known to good relative accuracy to apply the density of states method.
However, the strong residual sign problem means that the precise value of $\eitheta$ may not be computable if the fit parameter $a$ is close to $a_0$, and moreover, small uncertainties on $a$ will lead to  relative uncertainties on $\eitheta$ which can be of several orders of magnitude. 

As endnote, we observe that, although we have only presented results for $\sigma=4.2$, the strong residual sign problem is a general feature of the computation of $\eitheta$ over the distribution \eqref{fit} for any $\sigma$ and $a\to a_0(\sigma)$. As $\sigma$ is increased, which corresponds to larger volumes of the simulated physical system, the distribution \textit{looks} more and more normal and the Binder cumulant goes to three, however, the strong residual sign problem remains unmoved. A Gaussian value for the Binder cumulant is thus no guarantee for a Gaussian result for $\eitheta$. This will be the topic of a forthcoming paper discussing the phase distribution in a random matrix model of QCD, and its volume dependence \cite{Bloch:2018}.

\begin{acknowledgments}
I would like to thank Falk Bruckmann, Robert Lohmayer and Simon Reiser for useful discussions. This work was supported by the DFG collaborative research center SFB/TRR-55.
\end{acknowledgments}

\bibliography{biblio} 

\end{document}